\documentclass[%
reprint,
superscriptaddress,
 amsmath,amssymb,
 aps,
prb,
]{revtex4-2}

\usepackage{graphicx}%
\usepackage{dcolumn}%
\usepackage{bm}%
\usepackage{hyperref}%
\usepackage[utf8]{inputenc} %

\usepackage{xcolor}   %
\PassOptionsToPackage{english}{babel}
\DeclareUnicodeCharacter{03B4}{$\delta$}

\begin{document}

\title{Simulations of angle- and spatially-resolved vibrational electron energy loss spectroscopy for a system with a planar defect}

\author{Paul M. Zeiger}
\email{paul.zeiger@physics.uu.se}
\affiliation{%
  Department of Physics and Astronomy, Uppsala University, P.O. Box 516, 75120 Uppsala, Sweden
}%
\author{J\'{a}n Rusz}%
\affiliation{%
  Department of Physics and Astronomy, Uppsala University, P.O. Box 516, 75120 Uppsala, Sweden
}%

\date{\today}%

\begin{abstract}
Recent developments in experiments with vibrational electron energy loss spectroscopy (EELS) have revealed spectral shape variations at spatial resolutions down to sub-atomic scale. Interpretation in terms of local phonon density of states enables their qualitative understanding, yet a more detailed analysis is calling for advances in theoretical methods.
In Zeiger and Rusz, Phys. Rev. Lett. \textbf{124}, 025501 (2020) we have presented a frequency resolved frozen phonon multislice method for simulations of vibrational EELS. Detailed simulations for a plane wave electron beam scattering on a vibrating hexagonal boron nitride are presented in a companion manuscript (Zeiger and Rusz, arXiv:2104.03197). Here we present simulations of vibrational EELS assuming a convergent electron probe of nanometer size and atomic size on a hexagonal boron nitride structure model with a planar defect. With a nanometer beam we observe spectral shape modifications in the presence of the defect, which are correlated with local changes of the phonon density of states. With an atomic size electron beam, we observe the same, although with better contrast. In addition, we observe atomic level contrast and sub-atomic scale spectral shape modifications, which are particularly strong for small detector collection angles.
\end{abstract}

\maketitle

\section{Introduction}\label{sec:background}

Heat management is a major design limitation of integrated circuits today \cite{ogrenci-memik_heat_2015,theis_its_2010} and Moore's law is at the same time reaching its limits \cite{li_how_2019}. Advances in the field of phononic and thermoelectric materials allow for the control of phonons over large frequency regions and make it thereby possible to control the flow of heat \cite{maldovan_sound_2013}. Understanding the flow of heat at the nanometer and sub-nanometer level might lead to further progress in these fields and a technique, which can deliver this spatial resolution, is a prerequisite for such developments.

The (scanning) transmission electron microscope [(S)TEM] routinely allows to reach sub-{\AA} spatial resolution \cite{haider_electron_1998,batson_sub-angstrom_2002,akashi_aberration_2015,sawada_resolving_2015,morishita_resolution_2016,morishita_resolution_2018} and thanks to recent advances in monochromators it offers an energy resolution of electron energy loss spectroscopy (EELS) down to 4.2~meV \cite{krivanek_vibrational_2014,krivanek_progress_2019}. Unprecedented experiments such as mapping of bulk and surface modes of nanocubes \cite{lagos_mapping_2017}, investigations of the nature of polariton modes in Van der Waals crystals \cite{govyadinov_probing_2017_s}, temperature measurement at the nano-scale \cite{idrobo_temperature_2018,lagos_thermometry_2018}, identification and mapping of isotopically labeled molecules \cite{hachtel_identification_2019}, position- and momentum-resolved mapping of phonon modes \cite{hage_nanoscale_2018_s,senga_position_2019,venkatraman_simultaneous_2020}, atomic resolution phonon spectroscopy \cite{hage_phonon_2019,venkatraman_vibrational_2019}, functional group mapping \cite{collins_functional_2020}, single stacking fault \cite{yan_stacking_fault_2021}, and single-atom vibrational spectroscopy \cite{hage_single-atom_2020} were enabled by these advances in STEM instrumentation.

Venkatraman et al.\ have observed sub-atomic scale spectral changes in a silicon (Si) crystal, attributing them to selective phonon excitations as a function of electron beam impact parameter \cite{venkatraman_vibrational_2019}. In another work, Hage et al.\ have shown on a monolayer of graphene containing a single Si impurity atom, that vibrational STEM-EELS spectra show atomic scale changes near the Si-impurity \cite{hage_single-atom_2020}. Both works recognized the necessity of a general simulation method to address the complexity of observed phenomena, such that would be capable to treat arbitrary three-dimensional structures, including eventual defects, and allow for general beam and detector geometry. Contemporaneously we have reported on a frequency-resolved frozen phonon multislice method (FRFPMS) allowing efficient simulations of vibrational spectra, fully considering dynamical diffraction effects as well as beam and detector setups \cite{zeiger_efficient_2020}. In Ref.~\cite{zeiger_plane_wave_2021} we analyze the method in more detail and present results of calculations for a plane-wave electron beam. Recently, Yan et al.\ have presented space- and angle-resolved measurements of vibrational EELS modifications in presence of a stacking fault in silicon carbide, which were correlated with the local phonon density of states, however effects of electron beam propagation and scattering have not been considered \cite{yan_stacking_fault_2021}.

Here we present detailed simulations of space- and angle-resolved vibrational EELS on a model system of hexagonal boron nitride (h-BN) with AA' stacking. Two structure models are being considered. The first one is a periodic crystal without any structural defects \cite{zeiger_plane_wave_2021}. The second model contains a planar defect (an anti-phase boundary; APB) and the defect plane is oriented parallel to the electron beam. While such system is qualitatively similar to a stacking fault studied by Yan et al.~\cite{yan_stacking_fault_2021}, we note that here every atomic column consists of the same number of B and N elements and, as such, it is \emph{a priori} not likely to be revealed by high-angle annular dark field (HAADF) imaging. We show that the FRFPMS method resolves both angle-resolved spectral shape variations on a nanometer scale as well as sub-atomic scale spectral changes analogous to those reported in previous experimental works. Using a large off-axis detector geometry offers a clear detection of the defect, both with nanometer size probe and atomic size probe. The latter reveals minor sub-atomic scale spectral shape variations. A nanometer-sized electron probe gives furthermore access to the local phonon bandstructure using either small collection angles or $(q,E)$-mapping \cite{plotkin-swing_hybrid_2020,zeiger_plane_wave_2021}. %

\section{Methods}\label{sec:methods}

Most of the computational methods and parameters used here follow the choices made in the companion manuscript, where we describe the FRFPMS method in detail and consider the case of parallel illumination \cite{zeiger_plane_wave_2021}. For completeness, here we briefly summarize the methods and parameter settings, highlighting the differences in methodology and parameters between both works.

\begin{figure}
    \includegraphics[width=\linewidth]{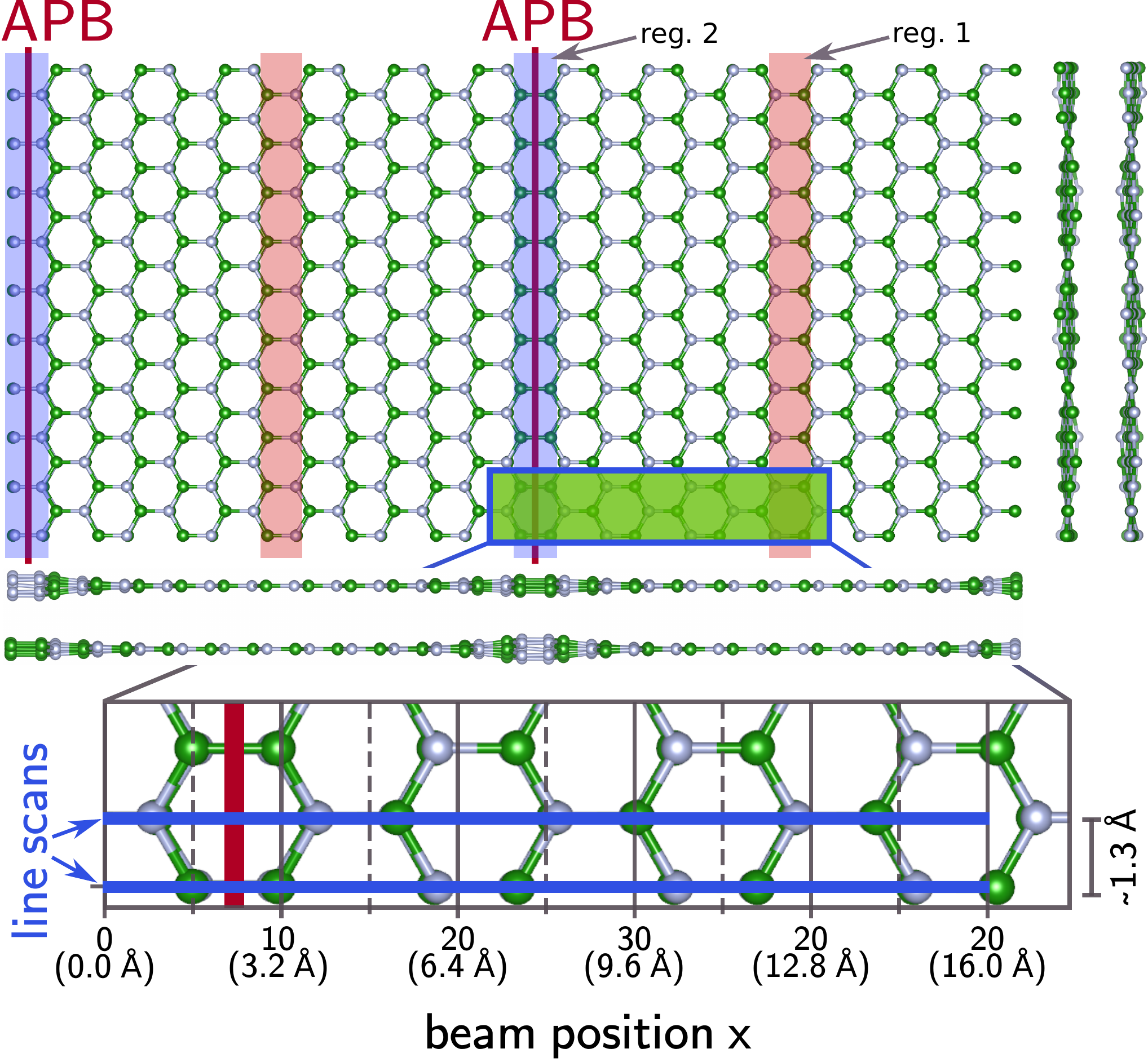}
    \caption{%
    Relaxed hBN structure with an antiphase boundary. The supercell viewed along the $z$-direction is displayed together with two side views of the two upper layers. Two types of areas are highlighted in red (region 1) and blue color (region 2), referring to an ``in grain'' region and APB region, respectively. The area scanned with atomic size electron beam (green rectangle) and line scans (thick blue lines) used in the subsequent analysis are highlighted in the bottom panel.}
    \label{fig:structure}
\end{figure}

We have performed independent simulations of an APB-free bulk hBN structure model and of a structure model containing two grains of bulk hBN and an APB at both interfaces between the grains (we apply periodic boundary conditions). Both structure models consist of the same total number of atoms ($N_\text{at} = 22080$) and hBN-layers ($N_\text{lay}=46$) in AA' stacking order. Our APB model can be obtained by slicing the pristine hBN structure once perpendicular to the hBN layers (here the $z$-direction) and shifting then one of the two obtained ``halves'' by one interlayer distance along the $z$-direction. As a result, the APB plane consists solely of B-B and N-N bonds. Electronic structure calculations show that such an APB is metastable in monolayer hBN \cite{liu_dislocations_2012,gomes_stability_2013}. %
The relaxed orthogonal simulation boxes have sizes of 52.01~{\AA} $\times$ 25.02~{\AA} $\times$ 149.71~{\AA} and 52.08~{\AA} $\times$ 25.13~{\AA} $\times$ 148.65~{\AA} for models without and with APB, respectively, and periodic boundary conditions were enforced in all three spatial directions. The structurally relaxed APB model, highlighting the atomic displacements in the neighborhood of the APB in the side-views, is visualized in Fig.~\ref{fig:structure}.

Structural relaxation as well as molecular dynamics calculations were performed with the \textsc{Lammps} software \cite{lammpsweb}. We have employed an extended Tersoff potential developed by Los et al.\ \cite{los_extended_2017} for intralayer interactions and for interlayer interactions we have used a potential of Ouyang et al., which was specifically optimized for the description of bulk hBN \cite{ouyang_mechanical_2020}. In addition, a shielded Coulomb potential accounts for partial charges on atoms between different layers \cite{maaravi_interlayer_2017}.

Phonon calculations (not shown), which were performed using a combination of the \textsc{phonolammps} \cite{phonolammpsgithub} and \textsc{phonopy} software packages \cite{thephonopygithub,togo_first_2015}, showed no imaginary frequencies for the structurally relaxed APB model. Furthermore, classical molecular dynamics simulations at constant pressure of 0.0~bar and for temperatures of up to 1200~K showed no signs of structural reconfiguration of the APB, confirming the stability of the structure. %

Non-equilibrium molecular dynamics calculations using the generalized Langevin equation have been done within \textsc{Lammps} using a hotspot thermostat \cite{dettori_simulating_2017,gle4md_web} with a width of $\Delta \omega = 2.5$~THz. Energy bins range from from 11.5~THz up to 51.5~THz with a step of 2.5~THz. In relation to Ref.~\cite{zeiger_plane_wave_2021}, we use here the coarser of the two energy grids treated there. Nevertheless, it was shown that the effective energy resolution is in both cases close to 10~meV. Considering that in STEM calculations we need to analyze significantly more electron beam position, this choice is computationally more efficient. The base temperature $T_{\text{base}}$ for the white-noise thermostat was set to zero and its damping parameter to $1/\gamma_\text{base}=0.1$~ps. Temperature at the peak of the hotspot thermostat was set to $T_\text{max}=300$~K and the hotspot damping parameter to $1/\gamma=0.5$~ps. Time step in these simulations was set to 0.5~fs.

In total $n_{\text{snap}} = 225$ snapshots were generated per energy bin by the non-equilibrium molecular dynamics simulations, starting after 25~ps of equilibration, taking a snapshot every 1~ps until the total trajectory length of 0.25~ns. The lateral multislice grid consisted of $1008 \times 480$ points. We have used the automatic potential slicing option of \textsc{DrProbe} \cite{barthel_drprobe_2018}.

In all visualizations of vibrational EELS below we show spectra multiplied by the energy loss. This is motivated by observations made in Ref.~\cite{zeiger_plane_wave_2021} as well as by the first Born approximation expression for the inelastic scattering cross-section containing the factor $1/\omega_{\mathbf{q},n}$ in all transition matrix elements, where the $\mathbf{q},n$ index a specific phonon mode. Energy-multiplied spectra are visually more easily comparable to the (local) phonon density of states, which aids the discussion.

High angle annular dark field (HAADF) $Z$-contrast simulations were performed using the standard frozen phonon multislice method \cite{loane_thermal_1991} by averaging over 60 snapshots. In contrast to the FRFPMS method, these snapshots were generated using constant temperature and constant volume molecular dynamics simulations at 300~K using a combination of microcanonical ensemble with Langevin dynamics, otherwise using the same structure model and multislice parameters as for the FRFPMS simulations. %

\section{Results}\label{sec:results}

\subsection{Local phonon density of states}

\begin{figure}
    \centering
    \includegraphics[width=\linewidth]{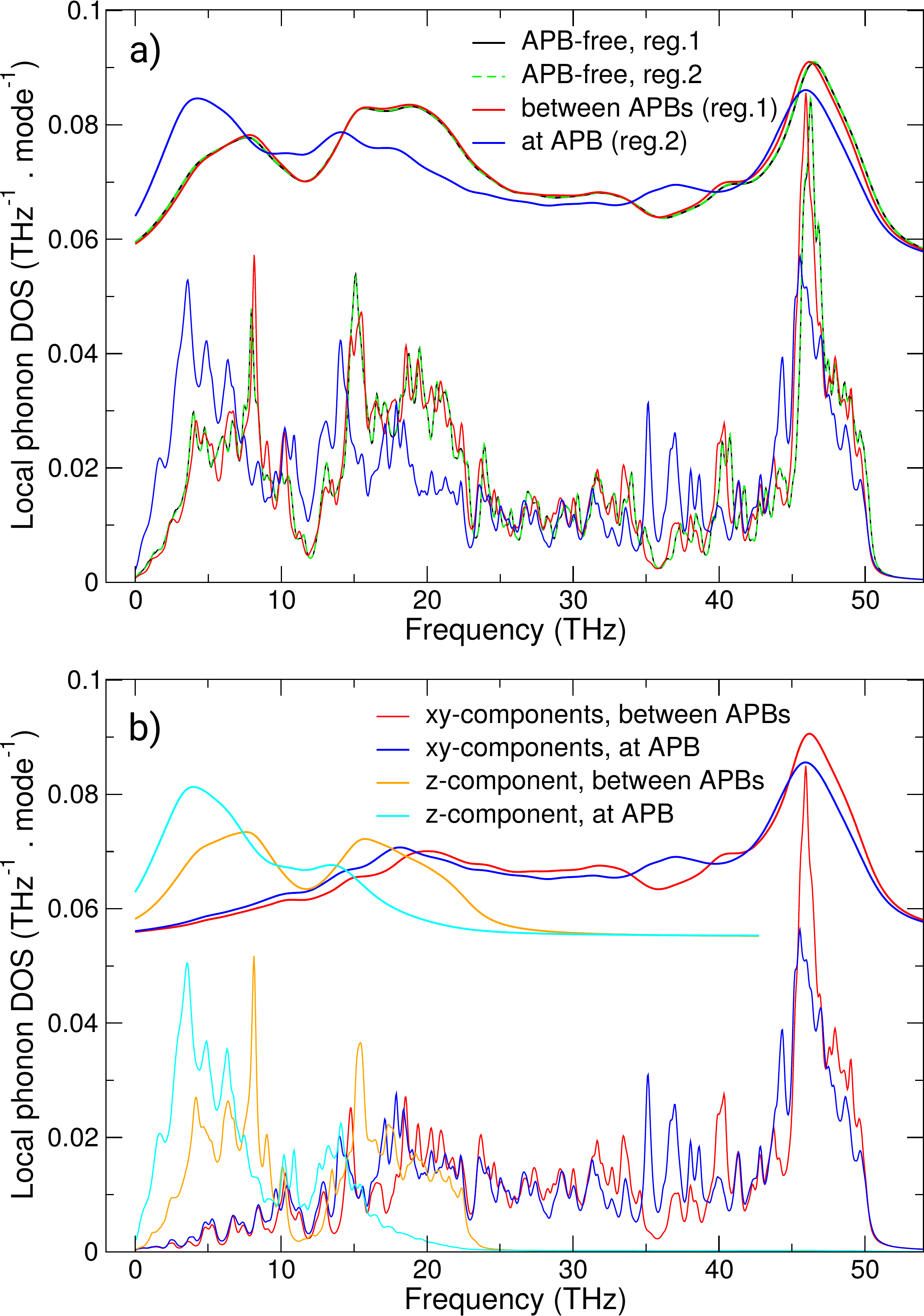}
    \caption{a) Comparison of local phonon densities of states calculated on a model with and without an antiphase boundary, respectively. b) Decomposition of the local phonon densities of states calculated for the model with antiphase boundary into their $z$-components (out of plane) and $xy$-components (in plane). In both panels, two local regions have been probed, see Fig.~\ref{fig:structure}. Lorentzian broadening of two different full-widths at half-maximum were used, namely 0.3~THz and 3.0~THz. The more broadened curves are vertically offset by $0.055$~THz$^{-1}$.mode$^{-1}$.}
    \label{fig:LPDOS}
\end{figure}

The top panel of Fig.~\ref{fig:LPDOS} shows a comparison of local phonon density of states (LPDOS) obtained for a system with and without defect. In Fig.~\ref{fig:structure} we have highlighted two areas for evaluation of the LPDOS. What is marked as \emph{region 1} is a set of atomic columns forming hexagons, that are positioned as far as possible from the two APBs. The \emph{region 2} then marks hexagons containing atoms forming the APB. For consistence and comparison purposes, we have defined the same areas within the APB-free structure model.

When using the structure model without defect, the LPDOS from both areas is expected to be the same and this is indeed the case, see the overlapping black and dashed green curves. The LPDOS is dominated by an optical peak at a frequency of around 46~THz and a set of less pronounced peaks at lower frequencies.

For a sufficiently large structure model, the LPDOS from \emph{region 1} in the APB structure model should match the LPDOS of the defect-free structure model. As Fig.~\ref{fig:LPDOS}a shows, the corresponding LPDOS (red curve) follows the LPDOS of the APB-free model very closely and we observe only a small ($\sim 0.2$ THz) redshift of the optical peak and a similar blueshift of acoustic modes. The broadened LPDOSs of both structure models are practically indistinguishable. Overall the agreement of the LPDOSs of APB and defect-free structure is excellent, suggesting that the structure model containing a defect is large enough to encompass regions where the impact of defects on the LPDOS is negligible.

As expected, the LPDOS of the defect region (blue curve) differs significantly from the other three LPDOSs. It has enhanced acoustic modes at low frequencies and the optical peak has a deformed shape, with a maximum redshifted by approximately 0.6~THz compared to the defect-free regions. Several other differences can be observed, for example an appearance of vibrational modes within the energy range of 35--39~THz, where the defect region LPDOS is much higher than in the defect-free region. Another pronounced difference is a reduction of the LPDOS in the defect region around 20~THz.

Figure~\ref{fig:LPDOS}b shows a decomposition of the LPDOS calculated for the structure model with defect into its components due to vibrations in the in-plane ($xy$) and out-of-plane ($z$) directions. Since the electron beam propagates along the $z$-direction, it is mostly sensitive to in-plane atomic vibrations \cite{dwyer_dipole_2018}. This will be important, when the calculated vibrational EELS spectra will be correlated with the LPDOS. For instance, between approximately 15 to 23~THz the total LPDOS is significantly larger in the defect-free region, however the decomposition into in-plane and out-of-plane components reveals that this is almost entirely due to vibrations in $z$-direction, which are expected to influence the electron beam only negligibly. The in-plane LPDOS in this energy region is of comparable amplitude in both regions, with the defect region LPDOS being on average somewhat higher.

\subsection{Vibrational EELS at nanometer scale}

We start our analysis with a nanometer-sized electron beam, to check whether the above-mentioned LPDOS differences can be detected in simulated vibrational EELS spectra. We have set up a calculation with a convergent probe of 3~mrad semi-angle, which leads to a beam full-width at half-maximum of approximately 0.9~nm at an acceleration voltage of 60~kV (neglecting any source-size broadening or aberrations). We have evaluated a grid of $12 \times 10$ beam positions, evenly spanning the whole structure model. That corresponds to a STEM sampling with steps of 0.43~nm along $x$-direction and 0.25~nm along $y$-direction, respectively. 

\begin{figure}
    \centering
    \includegraphics[width=\linewidth]{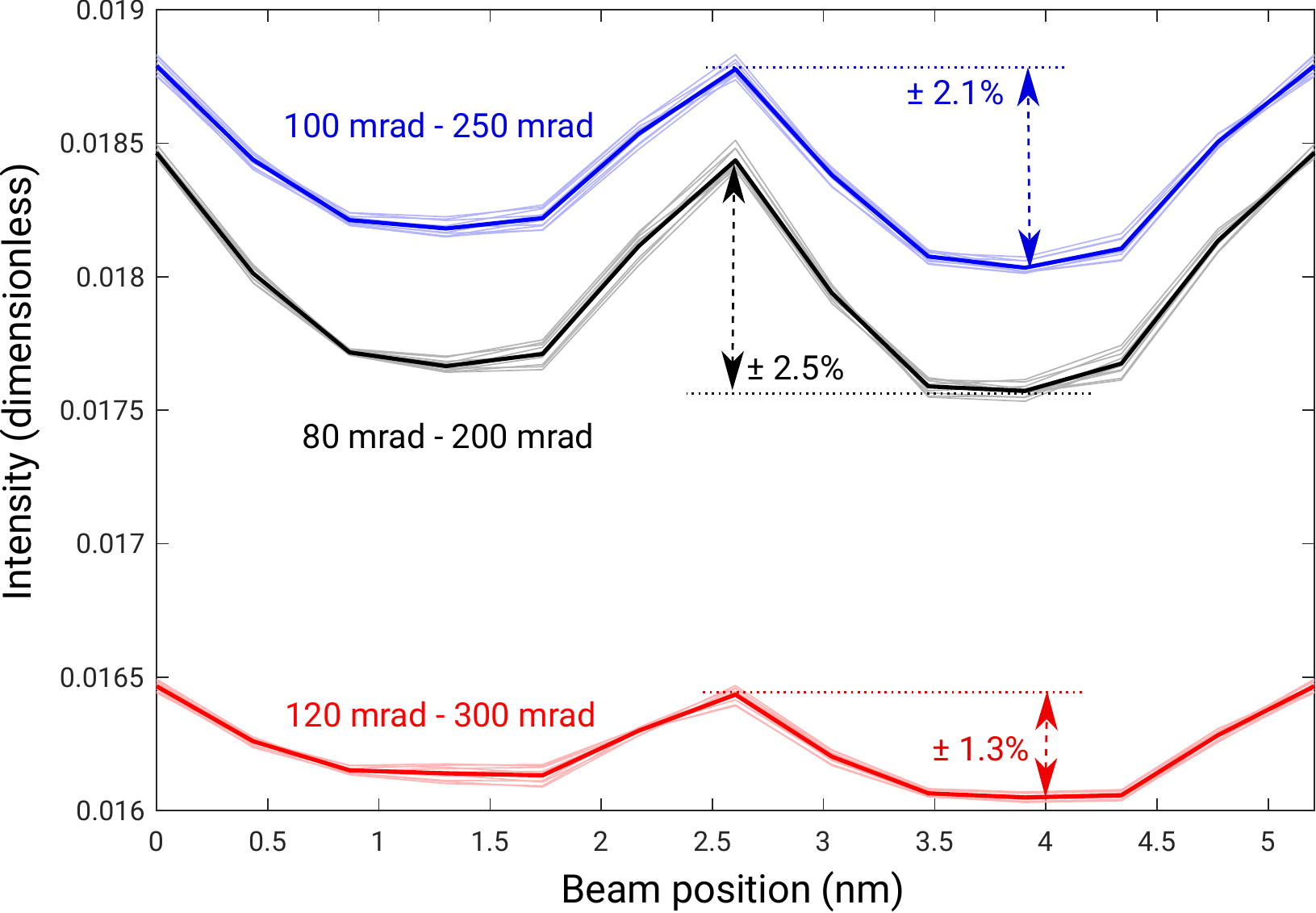}
    \caption{High-angle annular dark field profiles across the whole structure model with two defect planes. Ten individual profiles for different beam $y$-coordinates are plotted with thin lines of brighter colors. Their average is shown with thicker lines of saturated colors. Three different settings of inner and outer collection angles are indicated. The relative variation of intensity is given as the ratio of difference and sum of the maximal and the minimal intensity of each averaged profile.}
    \label{fig:haadf3mrad}
\end{figure}

Figure~\ref{fig:haadf3mrad} shows calculated HAADF profiles across the structure model with defect planes. By symmetry, we would not expect strong differences for beam positions differing by $y$-coordinate only, because the beam is too large for any atomic-scale contrast and the local differences in atomic positions and bonds are only along the $z$-direction---referring to the small displacement of atoms near the planar defect, seen in $xz$- or $yz$-projections of Fig.~\ref{fig:structure}. This is indeed the case, as can be seen in Fig.~\ref{fig:haadf3mrad}. Profiles shown with thick lines are averages over ten beam positions along the $y$-axis. Figure also includes the ten individual linear profiles with thin lines of a brighter color and they are obviously very close to each other and thus also to their average. Three different settings of inner and outer collection angles have been considered. Each of the profiles displays local maxima at the positions of defect planes. The HAADF intensity variation is, however, relatively weak. Defining it as a ratio of the maximal minus the minimal profile intensity and of their sum, the intensity variation ranges between 1.3\% to 2.5\%, i.e., it is a relatively weak effect. Yet it is nonzero, despite the same elemental composition of every single atomic column within the structure model. %
In the following paragraphs we will address the question, whether vibrational EELS has the capability to highlight the presence of the defect planes.

\begin{figure}
    \centering
    \includegraphics[width=\linewidth]{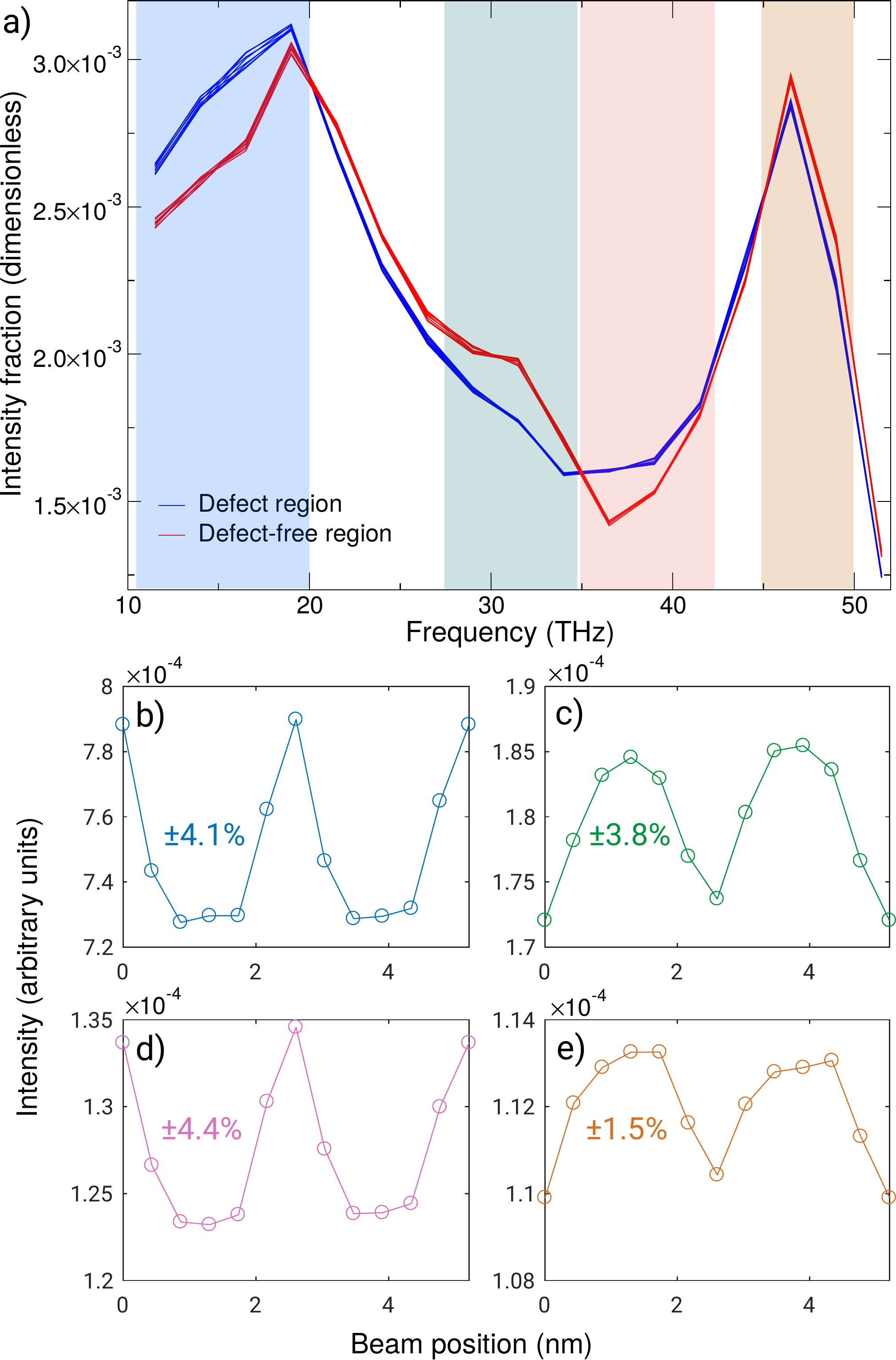}
    \caption{a) Vibrational EELS spectra calculated within the two regions defined in Fig.~\ref{fig:structure} assuming a nanometer-sized electron beam with convergence semi-angle of 3~mrad. Within each of the regions 10 spectra were evaluated at different beam $y$-coordinates. Four different frequency regions are marked by shading. Linear profiles across the whole structure model containing two planar defects are shown in panels b)--e) with frequency ranges and color coding corresponding to the shaded regions in panel a). Beam position step is approximately 0.43~nm resulting in 12 evenly spaced beam positions. Percentages refer to the intensity variations, defined as in Fig.~\ref{fig:haadf3mrad}.}
    \label{fig:spec3mrad}
\end{figure}

First we will analyze spectra calculated with a large off-axis detector. Specifically, we have set the detector center to (60~mrad,0) and its radius (collection semi-angle) to 25~mrad. In Fig.~\ref{fig:spec3mrad}a we show 10 spectra from the defect region (blue curves) and 10 spectra from the defect-free region (red curves). Within each region, the 10 individual spectra match closely each other, similarly as the individual HAADF profiles in Fig.~\ref{fig:haadf3mrad}. Mutual differences of individual spectra provide to a certain degree an estimation of the statistical error in our sampling. However, it should be pointed out that the 10 individual spectra are correlated to some degree, because the same structure snapshots have been used in their calculations and thus the same combinations of excited phonons are present. From a different perspective, atomic displacements in a given sample region do not unambiguously determine the displacements elsewhere in the sample, despite the reduced portfolio of available phonon modes within an energy bin, therefore the results suggest that the averaging over 225 snapshots provides a well converged result -- at least, for the selected large off-axis detector.

The difference of the spectra from the two regions is significantly stronger than the spread of the spectra within each region. This provides a theoretical support for the experimental observation that a nanometer sized electron beam is indeed sufficiently small to distinguish a planar defect \cite{yan_stacking_fault_2021}.

The spectral differences in Fig.~\ref{fig:spec3mrad}a match the qualitative differences between the LPDOS curves in Fig.~\ref{fig:LPDOS} well, such as the slight redshift of the optical phonon peak (barely seen at the energy resolution of 2.5~THz, which corresponds to approximately 10~meV), enhanced acoustic region (here specifically the region just above 10~THz) or the more intense defect-region spectrum between 35 and 40~THz, when compared to the spectrum from the defect-free region. Stronger spectrum from the defect-free region around 30~THz comes likely from the similar feature in LPDOS just above 30~THz. In relative terms, the modest enhancement of the acoustical modes within the defect region is more in line with the in-plane LPDOS component than with the total LPDOS (see Fig.~\ref{fig:LPDOS}), as was anticipated. It is somewhat surprising, that the intensity and shape of optical peaks come out rather similar in both spectra -- this is likely a consequence of relatively low energy resolution and 2.5~THz distance between energy bins.

A step further in mapping the nanometer scale spectral shape variations has been inspired by experiments of Yan et al.~\cite{yan_stacking_fault_2021}. As mentioned above, we have run spectral simulations for a grid of $12 \times 10$ beam positions evenly covering the whole structure model. The 12 evenly spaced beam positions correspond to a step of approximately $0.43$~nm in the direction perpendicular to the defect planes. Encouraged by very small spectral differences observed along the $y$-direction in Fig.~\ref{fig:spec3mrad}a, the 10 evenly spaced beam positions in $y$-direction have been averaged to improve statistics. In Fig.~\ref{fig:spec3mrad}a) we have highlighted four different energy ranges of interest, over which we have integrated the vertically averaged spectra and plotted them as intensity profiles across the structure model in $x$-direction. One can clearly observe the variation of the intensities correlating with the positions of the two planar defects in all highlighted energy ranges, an observation analogous to the enhancement of the acoustic modes in the vicinity of a stacking fault in SiC observed in Fig.~3 of Ref.~\cite{yan_stacking_fault_2021}. Note also that in comparison with the HAADF signal, the percentage of the intensity variation across the defect is increased in the vibrational spectra, ranging from $\pm 1.5$\% for optical modes up to more than $\pm 4$\% for both acoustic modes and frequency range 35--42~THz, respectively.

In order to get a deeper insight into the vibrational scattering of a nanometer-sized electron beam, we explore the information contained in spectra simulated with a small-collection angle detector. Specifically, we set the detector collection semi-angle to 3~mrad and we scan its center along specific lines within the diffraction plane. We focus on the region of scattering angles between 30~mrad and 110~mrad, 
where the inelastic (phonon) scattering represents a sizable fraction of the total scattering cross-section.

\begin{figure}
    \centering
    \includegraphics[width=\linewidth]{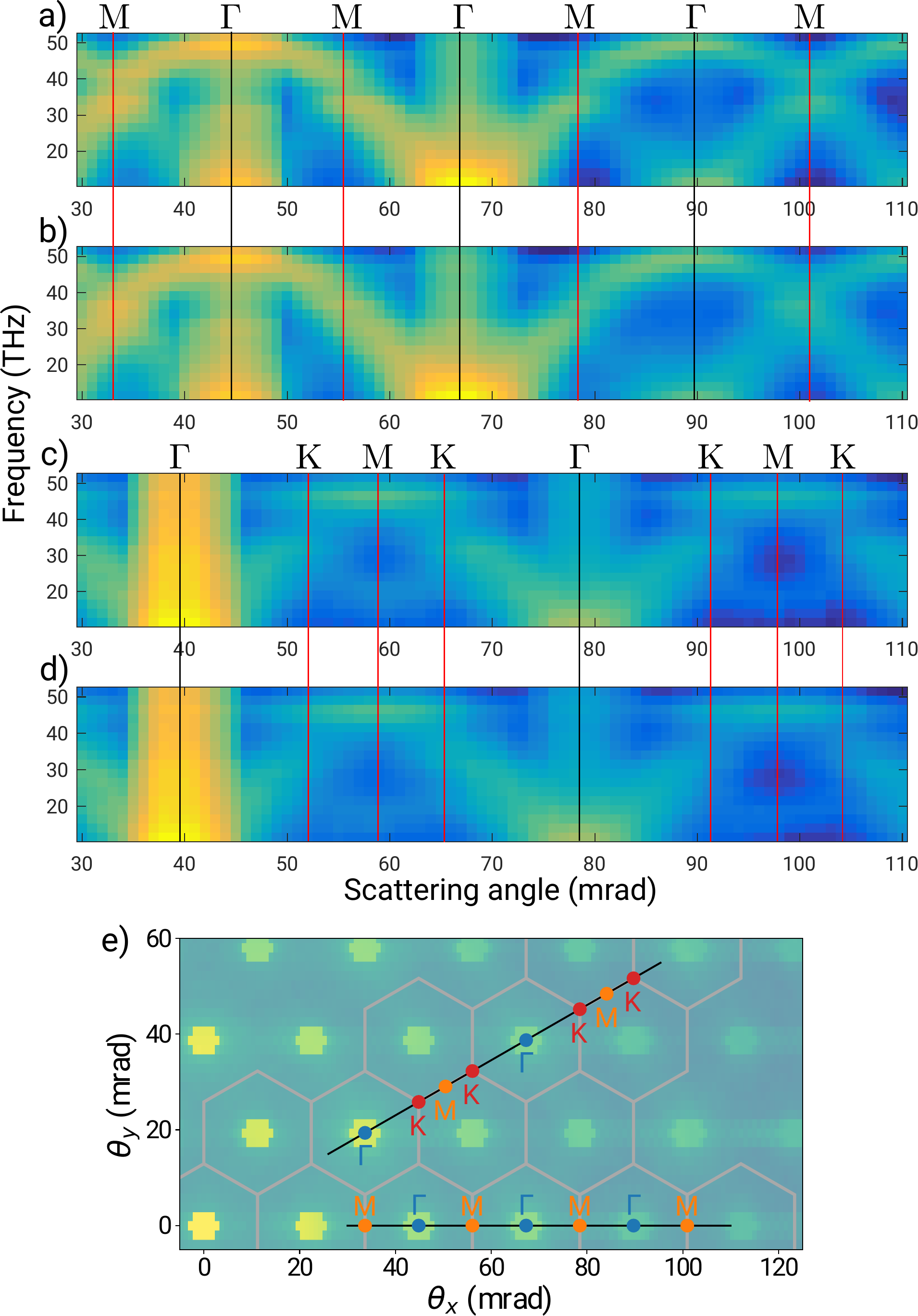}
    \caption{$(\mathbf{q},E)$ diagrams showing an evolution of local spectroscopic information as a function of scattering angle along selected directions in the $\mathbf{q}$-space. Panels a) and b) correspond to a scan along the $\theta_x$ scattering direction. Panels c) and d) correspond to a scan inclined under 30 degrees angle with respect to the $\theta_x$ direction. In each pair, the upper panel is calculated within the defect-free region and the lower one originates from the vicinity of the defect plane. The reciprocal space path of pure hBN is mapped on the scattering directions of hBN with defect and the vertical red lines mark the position of special points in the Brillouin zones. Subtle differences can be seen near $M$ and $K$ points and by generally higher intensity at lower energies within the defect region. Panel e) illustrates the $\mathbf{k}$-space paths visualized in panels a)--d) on top of a section of a diffraction pattern calculated by standard frozen phonon method for the nano-beam with a convergence semi-angle of 3~mrad.}
    \label{fig:qvsE3mrad}
\end{figure}

Figure~\ref{fig:qvsE3mrad} shows so called $(\mathbf{q},E)$-diagrams, which combine information about scattering intensity as a function of both energy and scattering angle (momentum transfer). Moreover, Fig.~\ref{fig:qvsE3mrad} also provides spatially resolved information, because panels a) and c) come from the defect-free region and panels b) and d) originate from the vicinity of the defect plane. Both the 3~mrad collection semi-angle and the 3mrad convergence semi-angle contribute to a blurring of the phonon bandstructure information contained in these images. Nevertheless, the information is still present and one can recognize dominant bands in the image. It is instructive to point out some asymmetries in the intensity of the phonon bands, for example in panel a) around $\Gamma$ at 90~mrad the V-shaped band due to acoustic modes is more intense towards higher scattering angles than it is at lower scattering angles. Also, the inverted U-shaped band corresponding to optical phonons is almost not present around $\Gamma$ at 67~mrad, while it is rather pronounced near 90~mrad. Similar features have been observed in works by Senga et al.~\cite{senga_position_2019} and in Plotkin-Swing et al.~\cite{plotkin-swing_hybrid_2020}. Note, in particular, the difference in spectra around $M$-points, when reached along the two distinct paths. Such difference is well expected, when one considers polarization vectors of the particular phonon branches (longitudinal or transversal) and their relative orientations to the propagation vector $\mathbf{q}$. Namely, the phonon scattering cross-section is proportional to a scalar product of $\mathbf{q}$ and polarization vector $\boldsymbol{\varepsilon}$ \cite{nicholls_theory_2019,dwyer_dipole_2018,senga_position_2019}. This is even more clearly seen for a plane-wave beam and point-like detection, see Ref.~\cite{zeiger_plane_wave_2021}.

Differences between the defect region, panels b) and d), and defect-free region, panels a) and c), seem relatively subtle. All the dominant features look very similar to each other, for both paths through the diffraction plane. A more detailed inspection reveals, however, differences near the Brillouin zone boundaries, i.e., in these plots nearby special points $K$ and $M$, respectively. Note, for example, how the non-crossing of the bands near $M$ point at 101~mrad gets blurred in the vicinity of the defect, Fig.~\ref{fig:qvsE3mrad}b). Also, at low frequencies, the minima around $M$-points and nearby $K$-points, respectively, are visibly deeper in the defect-free region, Fig.~\ref{fig:qvsE3mrad}a) and c), when compared to the defect-region counterparts, Fig.~\ref{fig:qvsE3mrad}b) and d). This well correlates with results shown in Fig.~\ref{fig:spec3mrad} and qualitatively with results of Ref.~\cite{yan_stacking_fault_2021}.

\begin{figure}
    \centering
    \includegraphics[width=0.85\columnwidth]{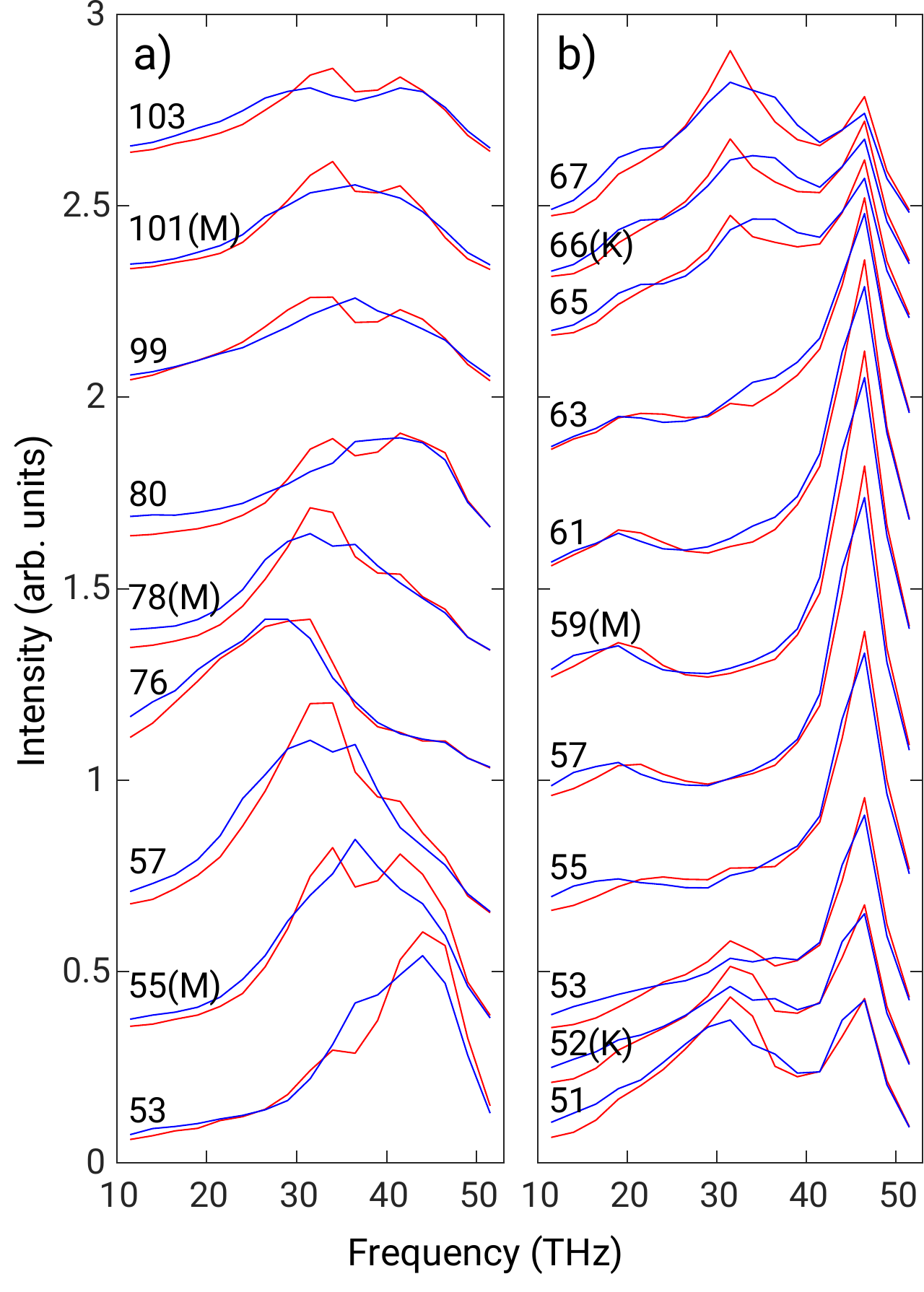}
    \caption{Vibrational EELS spectra calculated for selected scattering angles, averaged over 10 different $y$-coordinates of the beam position. Convergence and collection semi-angles are both set to 3~mrad. Red lines correspond to defect-free region and blue lines to the defect-region, respectively. Panel a) shows spectra along the horizontal path in the diffraction plane, as in Fig.~\ref{fig:qvsE3mrad}a,b. Panel b) shows spectra along a path inclined 30~degrees with respect to the horizontal axis, as in Fig.~\ref{fig:qvsE3mrad}c,d. Scattering angles in mrad along these paths, as well as the special points, are labeled.}
    \label{fig:spectra}
\end{figure}

These differences, along with other effects, get highlighted in a one-to-one comparison. Figure~\ref{fig:spectra} shows selected spectra from Fig.~\ref{fig:qvsE3mrad}, comparing the defect-free region (red) with the defect-region (blue). Several interesting observations can be made.

In Fig.~\ref{fig:spectra}a there are spectra from close neighborhoods of three different M points, see Fig.~\ref{fig:qvsE3mrad}a,b at 55~mrad, 78~mrad and 101~mrad, respectively. First, it is striking, how mutually different these three groups of spectra are. This is likely caused by dynamical diffraction, which excites individual Bragg discs with different strengths at a thickness of 15~nm. Without this elastic scattering effect, the intensity of inelastic scattering would decrease with increasing scattering angle due to the $1/q^2$ factor in the scattering cross-section. However, more intense Bragg spots may enhance inelastic scattering in their neighborhood by bringing the low-$q$ inelastic transitions to larger scattering angles. This is best seen in Fig.~\ref{fig:qvsE3mrad} around the (300) spot (scattering angle approx.\ 66~mrad), where optical modes are relatively weak, while the acoustic modes in its neighborhood are stronger than around the (200) spot (scattering angle approx.\ 44~mrad).

The second observation, which we would like to highlight here, is an apparent disappearance of the gap between the longitudinal acoustic (LA) and longitudinal optical (LO) modes between around 34 to 39~THz, c.f. Figs.~\ref{fig:LPDOS} and \ref{fig:spec3mrad}. Note how the spectra from the defect-free region (red) consist of two peaks approximately around 34~THz and 41~THz, corresponding to LA and LO modes, respectively. This double-peak feature is missing in the spectra from the defect-region (blue). Instead, the spectra from the defect region have usually a higher intensity at around 37~THz than their defect-free region counterparts, in some case even forming a clear peak there.

In Fig.~\ref{fig:spectra}b we highlight one of the K-M-K segments from the $\mathbf{q}$-space path shown Fig.~\ref{fig:qvsE3mrad}c,d. First, we will comment on the qualitative difference of the spectrum at M point from those in Fig.~\ref{fig:spectra}a. While we see only longitudinal modes at M points in panel a), in panel b) we instead see the transversal modes only. This can be qualitatively understood in terms of the different relative orientation of the $\mathbf{q}$-vector and the corresponding polarization vectors $\boldsymbol{\varepsilon}$ -- see a discussion and phonon band-structure, Fig.~1, in Ref.~\cite{zeiger_plane_wave_2021}. In panel b) the transversal optical (TO) mode leads to a sharp peak in the 46.5~THz energy bin and the transversal acoustic (TA) mode shows up as a weaker peak around 20~THz.

\begin{figure}
    \centering
    \includegraphics[width=\linewidth]{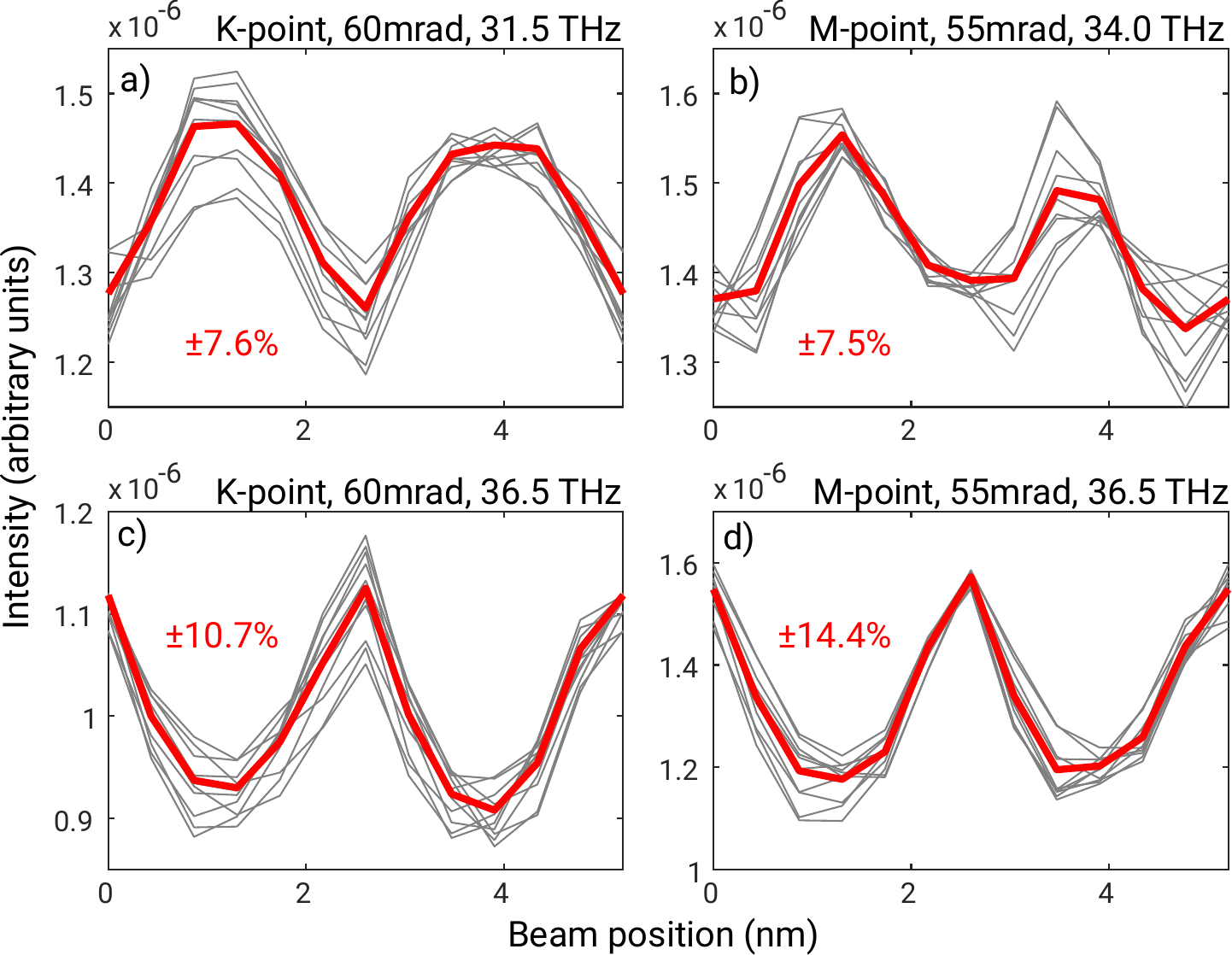}
    \caption{Profiles of the vibrational EELS intensity across the structure model with planar defects at selected special points in the reciprocal space at specific values of energy losses. Panels a) and c) refer to a K-point located 66~mrad from the direct beam along the line 30-degrees tilted from $k_x$-axis, see Fig.~\ref{fig:qvsE3mrad}c,d. Panels b) and d) refer to an M-point located 55~mrad from the direct beam along the $k_x$-axis, see Fig.~\ref{fig:qvsE3mrad}a,b. Energy loss is indicated by the frequency bin in the top right corner of every panel. Gray lines show 10 individual profiles for varying $y$-coordinates of the scan and thick red line shows an averaged profile. Percentages refer to intensity spread of the averaged profiles, as defined in Fig.~\ref{fig:haadf3mrad}.}
    \label{fig:KandMprofile}
\end{figure}

The second observation is the difference of spectral intensities around the K points, when comparing the spectra from defect-free region and the defect region, especially in the energy range between 30--40~THz. This correlates well with the similar qualitative feature in Fig.~\ref{fig:spec3mrad}a calculated for a large off-axis detector, however the flip of intensities seems to be enhanced in the surrounding of the K points.

We conclude this subsection by discussing profiles of the spectra at selected scattering angles and energy losses, where we identified in Fig.~\ref{fig:spectra} notable differences between the defect and defect-free regions, respectively. Figure~\ref{fig:KandMprofile} show profiles for selected K- and M-points across the whole structure model. Ten profiles from individual scan lines (differing by $y$-coordinate) are shown in gray color and their average is shown in red color. Here we observe a much larger spread of values, when compared to the spectra calculated for large off-axis detector in Fig.~\ref{fig:spec3mrad}. This is well expected due to significantly smaller detector area (3~mrad vs 25~mrad), while detector centers reside at similar scattering angles (55~mrad and 60~mrad, respectively). By averaging over a larger number of structure snapshots the spread would likely decrease. Nevertheless, the averaged profiles (red) show clear structure revealing the position of the defect planes. The intensity variation in these profiles is much larger than in Fig.~\ref{fig:spec3mrad}, reaching about $\pm 14$\%.
That leads to a proposal that experiments with nanobeam aiming to detect an antiphase boundary in hBN would be most likely to succeed when measuring spectra at $M$-points along the horizontal path in the diffraction plane and at $K$-points along the 30-degree-inclined path. We note that at higher energy resolution, both in theory as well as in experiment, the spectral differences are likely to be even more pronounced than presented here.

\subsection{Vibrational EELS at atomic scale}

\begin{figure}
    \centering
    \includegraphics[width=\columnwidth]{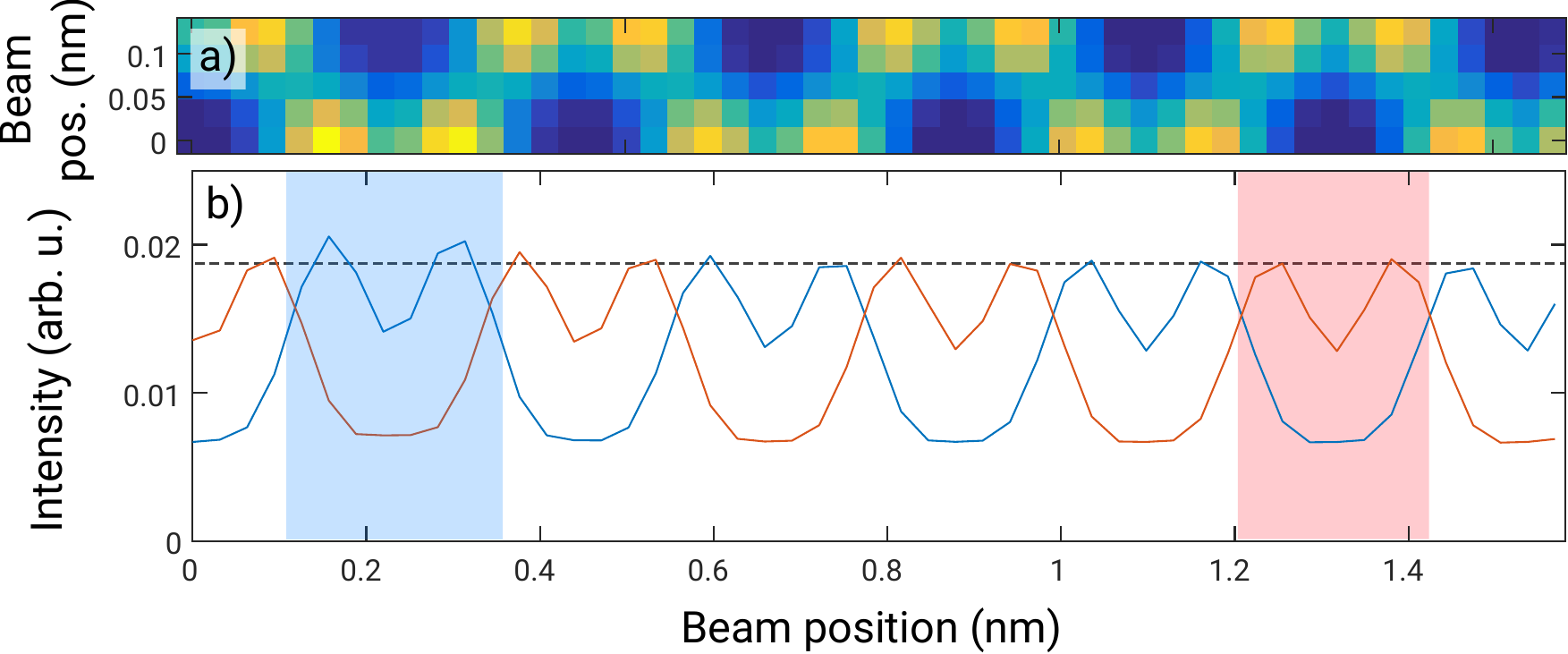}
    \caption{a) High-angle annular dark field image within the shaded area of the structure model, Fig.~\ref{fig:structure}. b) Line profiles from the HAADF image passing through the bottom (blue) and top (red) row of the atomic columns. Horizontal dashed line traces intensity peaks at the positions of atomic columns in defect-free region. Only the atomic columns directly involved in the defect plane deviate from this profile. Shaded areas are used to extract averaged spectra below.}
    \label{fig:haadf30mrad}
\end{figure}

We proceed with simulations involving atomic size electron beam with convergence semi-angle of 25~mrad. Neglecting any aberrations, at 60~kV acceleration voltage, this results in a probe size of approximately 1.1~\AA{}. In Fig.~\ref{fig:structure} we have highlighted an area of size approximately 1.57~nm $\times$ 0.125~nm (green-shaded and magnified region), which was scanned by a grid of $51 \times 5$ beam positions with a grid step of approximately 0.31~\AA{} in both $x$- and $y$-directions.

Analogous to the previous section, we start our analysis with an HAADF image, assuming inner and outer collection semi-angles of 120 and 300~mrad, respectively, see Fig.~\ref{fig:haadf30mrad}. In panel a) we show a common atomic-resolution STEM-HAADF image. The atomic columns are clearly resolved. A careful eye might notice a slightly higher intensity of the two left-most atomic columns in the bottom row. Those are the columns directly involved in the APB, forming B--B and N--N bonds. This enhancement of intensity is more clearly seen in linear profiles shown in panel b). The blue and red profiles correspond to the bottom and top lines of the HAADF image, respectively. The dotted line, which serves as a guide for the eye shows that the intensity of almost all atomic columns is remarkably similar. Exceptions are the atomic columns directly involved in the B--B and N--N bonds, which are approximately 10\% more intense, and their nearest neighbors in the upper line of the HAADF image, which are about 3\% more intense. All the other atomic columns have the same intensity within the errors of the sampling (both the number of snapshots and the step size of the grid of beam positions).

These findings from Fig.~\ref{fig:haadf30mrad} suggest, that the defect is very localized, notably influencing only those atomic columns that are directly involved in the defect plane. The intensity change is stronger than what we have seen in Fig.~\ref{fig:haadf3mrad} with a nanometer-sized beam. This could be expected due to an order of magnitude of difference in electron beam diameters. A nanometer-sized electron beam centered on the defect plane hits also atomic columns several \AA{}ngstr\"{o}ms away from the defect plane, partly blurring the local information. Thus, we would expect to see also stronger spectral changes between the defect-region and the defect-free region. For that purpose we have highlighted two areas in Fig.~\ref{fig:haadf30mrad}b, from which we extract averaged spectra below.

Note that Yan et al.~\cite{yan_stacking_fault_2021} in their Extended Data Fig.~1 also observe a similar intensity enhancement at atomic scale, although their raw line profile shows a qualitatively different behavior -- a locally reduced intensity oscillation with a positive offset in the vicinity of the stacking fault. We instead observe the same baseline intensity in between atomic columns and locally increased intensity in the HAADF image. This could be explained by a lower degree of alignment of atoms within atomic columns parallel to the beam direction in the vicinity of the observed SiC stacking fault, likely caused by a gradual relaxation of strain introduced by lattice mismatch of Si substrate and SiC grown on it.

\begin{figure}
    \centering
    \includegraphics[width=\columnwidth]{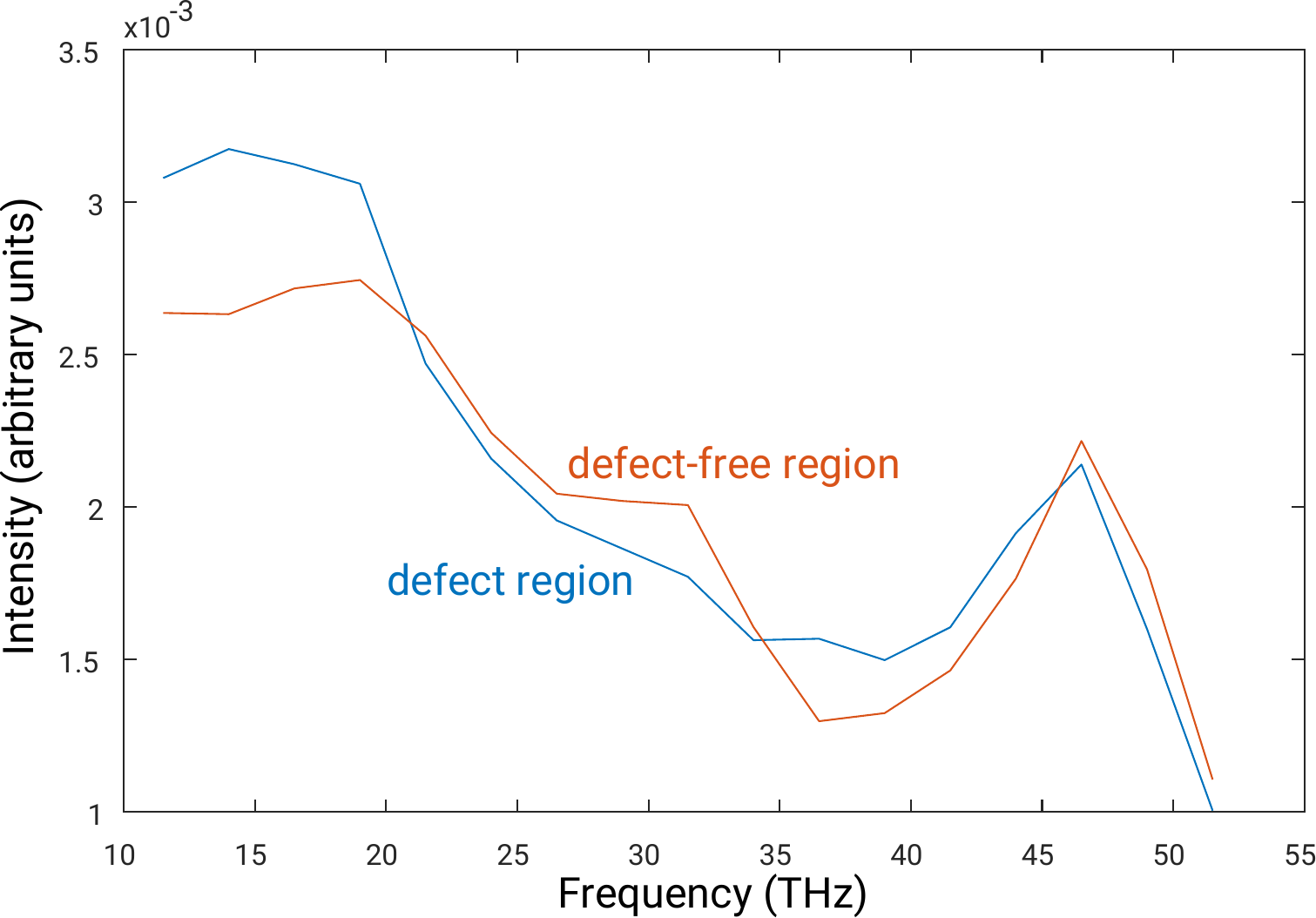}
    \caption{Averaged spectra evaluated within the shaded regions in Fig.~\ref{fig:spec30mrad}. Blue and red spectra correspond to defect and defect-free regions, respectively.}
    \label{fig:spec30mrad}
\end{figure}

Before proceeding to the discussion of the spectra, we note again that every atomic column in our structure model consists of the same number of B and N atoms, whether it is directly within the defect planes, or near or far from them. Seeing thus 10\% differences in the intensity among these columns goes against the interpretation of the atomic resolution HAADF images as a function of composition of the columns. Note that this effect would likely not be seen within an Einstein model treatment of atomic vibrations, because it is arguably the local difference of the atomic vibration modes (an overall redshift, see Fig.~\ref{fig:LPDOS}) due to the presence of the defect. In that sense, the scenario presented here differs from the stacking fault in Yan et al.~\cite{yan_stacking_fault_2021}, where an atomic resolution HAADF image clearly reveals displaced atomic columns. Here there is no such information available in the HAADF image alone and the locally increased intensity around the APB, without performing vibrational EELS, could be easily misinterpreted as a locally thicker sample. This finding may have consequences for methods related to atom counting \cite{debacker_atom_counting_2013}.

For each beam position, at which we have evaluated the HAADF intensity, we have also calculated a full $(q_x,q_y,E)$ datacube of vibrational spectra. Figure~\ref{fig:spec30mrad} shows vibrational EELS averaged from the shaded regions within Fig.~\ref{fig:haadf30mrad}b, evaluated for a large off-axis detector, similar to the experimental geometry of Hage et al.\ \cite{hage_single-atom_2020}, specifically, 25~mrad collection semi-angle displaced by 60~mrad along the $\theta_x$-direction. These spectra remind of Fig.~\ref{fig:spec3mrad} calculated with a nanometer-sized electron beam, nevertheless, as anticipated, the spectral shape differences are more pronounced here. Within the HAADF image we have seen that only the atomic columns directly involved in the defect plane show intensity differences. We can inspect, whether this is the case also for vibrational EELS, by plotting individual spectra along the scan lines, for which we have plotted the HAADF linear profiles in Fig.~\ref{fig:haadf30mrad}b.

\begin{figure}
    \centering
    \includegraphics[width=\columnwidth]{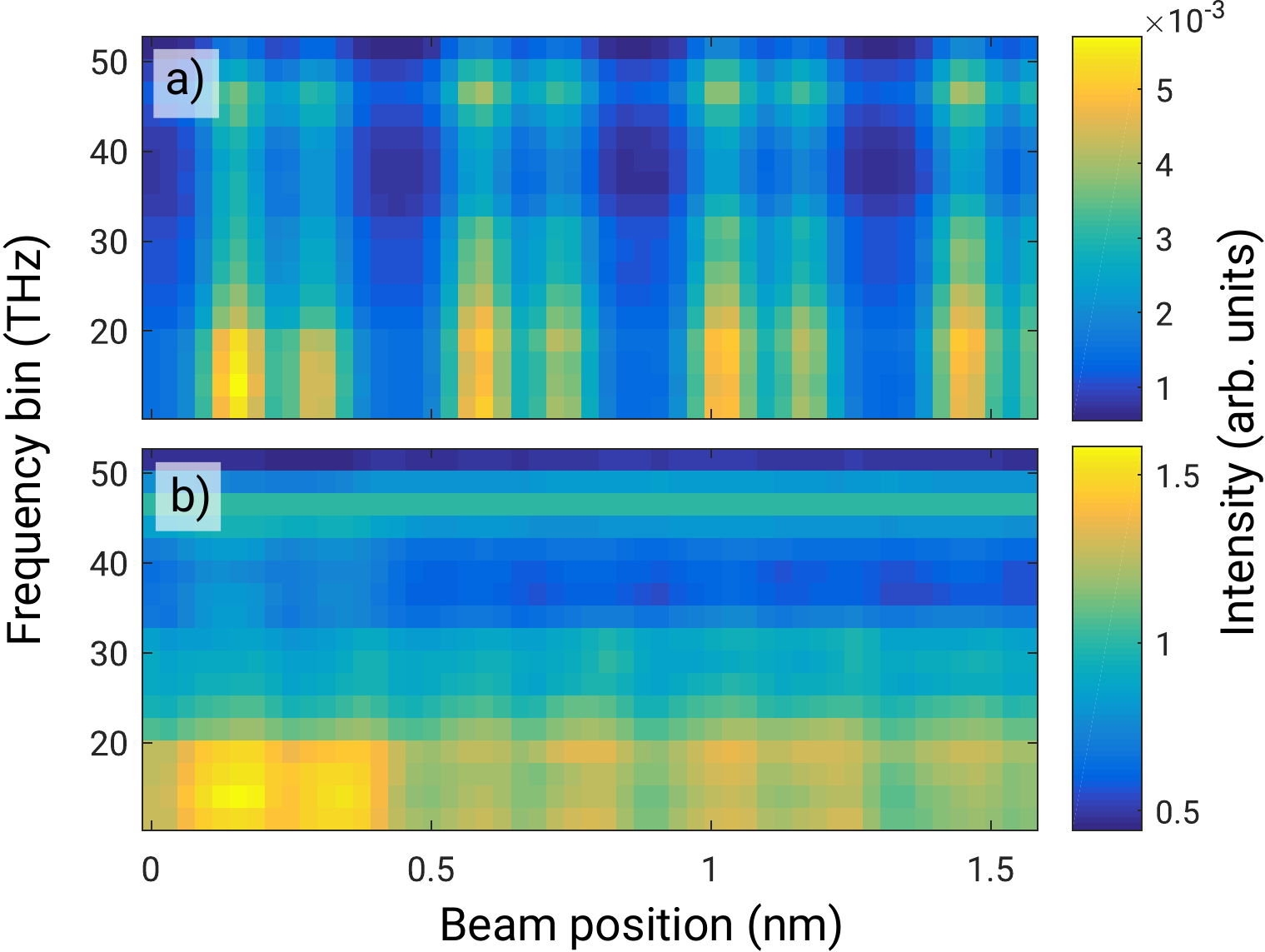}
    \caption{$(\mathbf{R},E)$-diagrams combining spectral information (vertical axis) and spatial dimension (horizontal axis). The horizontal axis corresponds to the bottom scan-line from Fig.~\ref{fig:structure}. Raw spectra are shown in panel a), while in the panel b) the same spectra are shown, just normalized to the same intensity at the optical peak (46.5~THz).}
    \label{fig:REdiag}
\end{figure}

Figure~\ref{fig:REdiag} shows two-dimensional datasets, where the vertical axis represents the spectral dimension (frequency / energy loss) and the horizontal axis is the real-space coordinate. It corresponds to a line scan, where the electron beam follows the bottom scan-line in Fig.~\ref{fig:structure}. This type of image can be understood as a real-space analogue of the $(\mathbf{q},E)$-diagrams shown in Fig.~\ref{fig:qvsE3mrad}. Two ways of visualizing the same data will aid the discussion below. First, we show in the top panel non-normalized raw spectra and in the bottom panel we show spectra normalized to the optical phonon peak in the 46.5~THz energy bin. %

Non-normalized spectra, Fig.~\ref{fig:REdiag}a, show a clear atomic scale contrast with maxima at the positions of atomic columns. This agrees with experiments and calculations in Hage et al.\ \cite{hage_phonon_2019} and with simulations in our previous work \cite{zeiger_efficient_2020}. We have verified that spectra calculated at beam positions about 12~\AA{} away from the APB plane closely follow spectra obtained from an APB-free model. Focusing on the spectra from the closest vicinity of atomic columns, there is a certain enhancement of the intensity of phonon peaks below $\sim 20$~THz when the beam passes through the APB plane and its closest vicinity. We will return to this enhancement in the next paragraph. Interestingly, the pairs of neighboring atomic columns have different intensity, the left one always appearing somewhat more intense than the right one. This is caused by an asymmetric placement of the detector aperture with respect to the rest of the beam-specimen system. We have verified that centering the off-axis detector around $(-60,0)$~mrad would lead to a swapping of column intensities within these pairs, making the right one more intense. The same behavior is actually also observed for pure hBN. One can observe some additional differences between the spectra from the APB region as compared to the defect-free region, nevertheless they are relatively subtle in this graphical representation.

\begin{figure}
    \centering
    \includegraphics[width=\columnwidth]{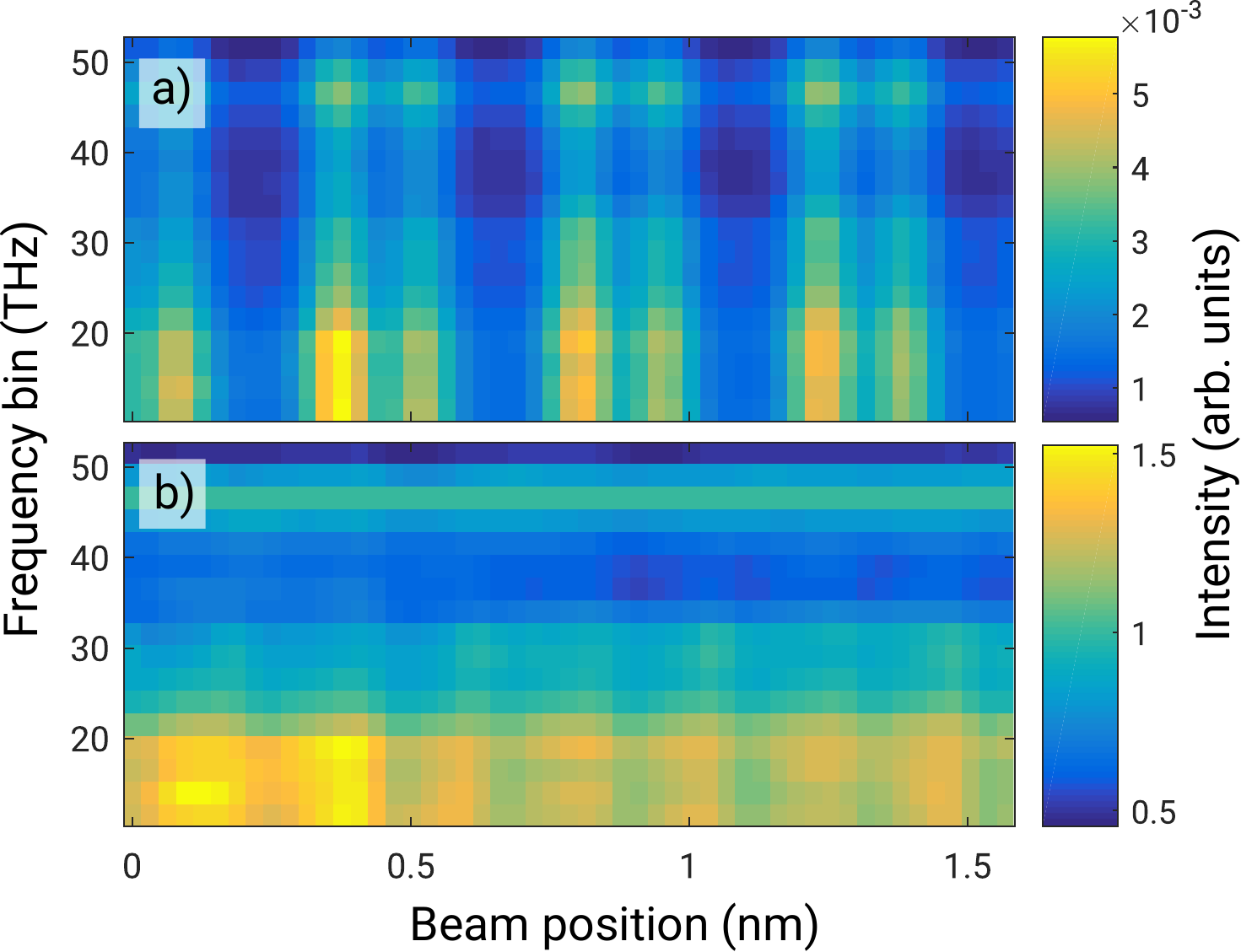}
    \caption{$(\mathbf{R},E)$-diagrams along the upper scan-line within the scanned area (see Fig.~\ref{fig:structure}). Raw spectra are shown in panel a), while in the panel b) the same spectra are shown, just normalized to the same intensity at the optical peak (46.5~THz).}
    \label{fig:REdiag5}
\end{figure}

The spectral signatures of the defect become more evident, when we normalize all the spectra to the optical peak at 46.5~THz, as is shown in Fig.~\ref{fig:REdiag}b. The spectral shape is distinctly different within the region of approximately $\pm 2$~\AA{} from the center of the APB plane. Apart from the above-mentioned enhancement below 20~THz, note also the enhancement in frequency bins between 34--44~THz. Beyond 2~\AA{} from the center of defect plane this local modification disappears and one can observe only slight spectral shape modifications correlated with on-column vs off-column beam positions. Note that these spectral shape modifications happen at sub-atomic scale. For instance, the relative intensity of the acoustic region with respect to the optical phonon peak is lower for beam positions in between the atomic columns. %

Figure~\ref{fig:REdiag5} presents $(\mathbf{R},E)$-diagrams for the beam positions along the upper line-scan highlighted in Fig.~\ref{fig:structure}. Qualitatively the findings are similar as in Fig.~\ref{fig:REdiag}: 1) there is the same asymmetry in the intensity of nearest-neighbor pairs of atomic columns, 2) sub-atomic scale spectral shape variation of the same kind, 3) enhancement of intensity within the acoustic region for beam positions within the hexagon containing the defect plane and, in comparison to Fig.~\ref{fig:REdiag}, a somewhat weaker enhancement of intensity within the frequency interval 34--44~THz, suggesting, that these modes could be highly localized on the atoms forming B--B and N--N bonds at the defect. Note, however, that the extent of the modification of the vibrational EELS is wider than in the case of the HAADF signal (Fig.~\ref{fig:haadf30mrad}). There the intensity changes were almost entirely limited to the atomic columns forming the B--B and N--N bonds, while in the case of the vibrational EELS, all atomic columns within the hexagons containing the defect plane lead to visibly modified spectra. Nevertheless, these spectral shape modifications due to the presence of the defect are still confined within a $\pm 2$~\AA{} interval around the defect plane.

The observed behavior qualitatively matches findings of Hage et al.\ \cite{hage_single-atom_2020} in terms of sub-nanometer scale modification of the vibrational spectrum in the vicinity of a localized defect and extends it with an observation of minor sub-atomic scale spectral shape variations, which are likely below the sensitivity of the reported experiment.

Sub-atomic scale spectral shape variations were, however, observed by Venkatraman et al.~\cite{venkatraman_vibrational_2019}. They have measured vibrational EELS spectra with an on-axis detector and observed substantial sub-atomic scale variations in vibrational EELS for an on-axis detector geometry and a collection semi-angle of 12~mrad. For a larger collection semi-angle of 24~mrad, however, the spectral shape appeared to be independent of the beam position within the experimental resolution and only variations in intensity were observed.

\begin{figure}
    \centering
    \includegraphics[width=\columnwidth]{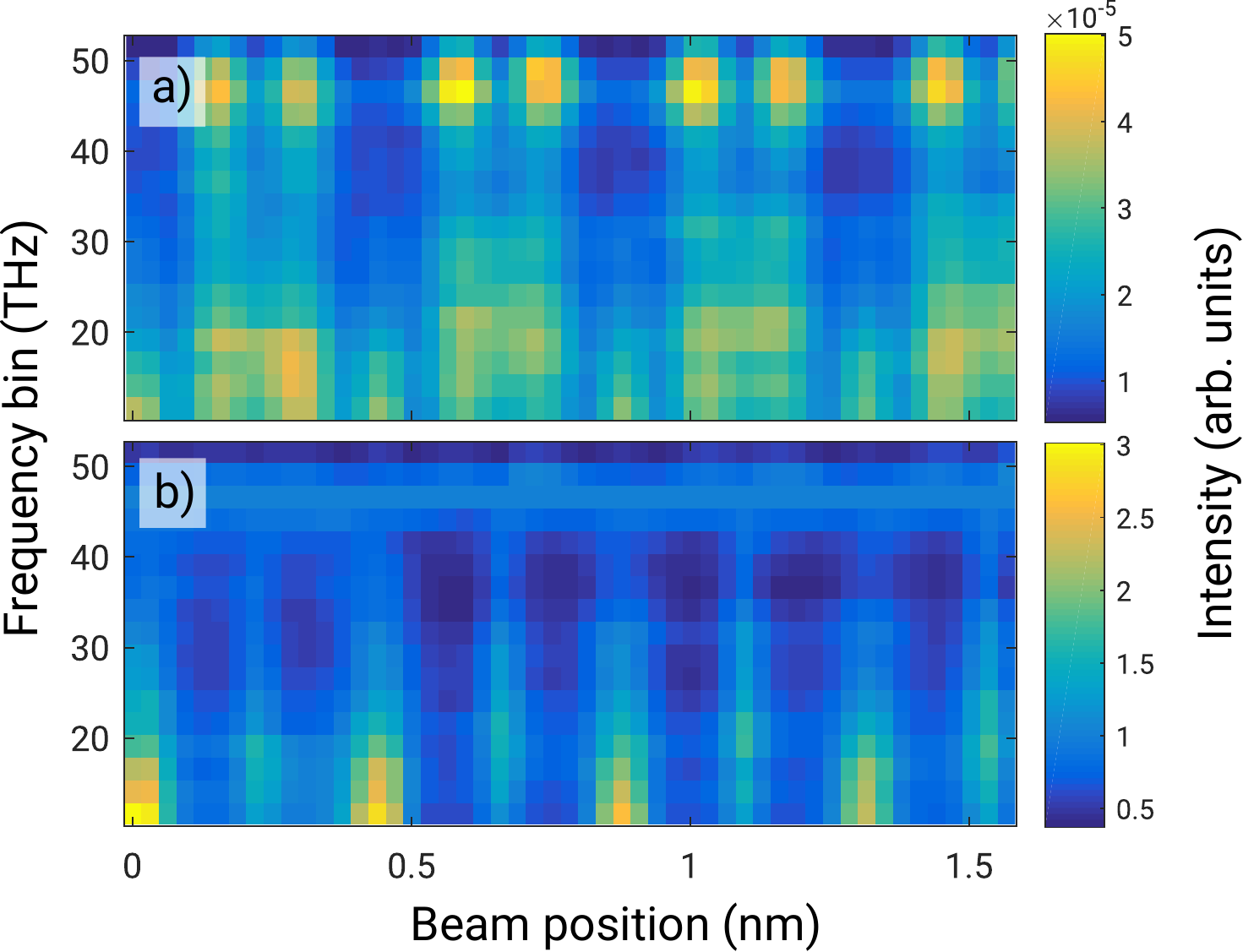}
    \caption{$(\mathbf{R},E)$-diagrams like in Fig.~\ref{fig:REdiag}, evaluated for a 3~mrad collection semi-angle centered around $\Gamma$-point at $\theta_x=67$~mrad.}
    \label{fig:REdiag3mrad}
\end{figure}

In its present form, our method doesn't include dipole scattering, therefore our calculations for on-axis geometry would not be realistic. Also, this is the area of scattering angles, where coherent component of the scattering cross-section dominates (see our discussion related to Fig.~\ref{fig:qvsE3mrad} and its range of scattering angles). Nevertheless, we can probe, whether a reduced collection semi-angle would enhance sub-atomic scale variations of the vibrational EELS.

In Fig.~\ref{fig:REdiag3mrad} we show $(\mathbf{R},E)$-diagrams for a collection semi-angle of 3~mrad centered around a $\Gamma$-point at scattering angle 67~mrad displaced from the center of the diffraction pattern along the $\theta_x$-axis. In panel a) with non-normalized spectra we note that in contrast with large-collection angle case, here the intensities at atomic columns within the defect region are reduced. Nevertheless, in the context of the previous paragraphs, it is very interesting to analyze panel b), where all spectra are normalized to the optical peak at 46.5~THz. We observe striking differences in spectral shape along the scan line. There is a notable enhancement of relative intensity at lower frequencies, whenever the electron beam is in between the atomic columns. In contrast, at atomic columns the optical peak is the dominant feature. In the defect region we observe an enhancement of the relative intensity between 34--44~THz, especially at the position of atomic columns. Note also the range of the associated colorbar, covering double of the range seen in Figs.~\ref{fig:REdiag} and \ref{fig:REdiag5}. Our findings thus corroborate the significant sub-atomic scale spectral shape variations at reduced collection semi-angles.

\begin{figure}
    \centering
    \includegraphics[width=\columnwidth]{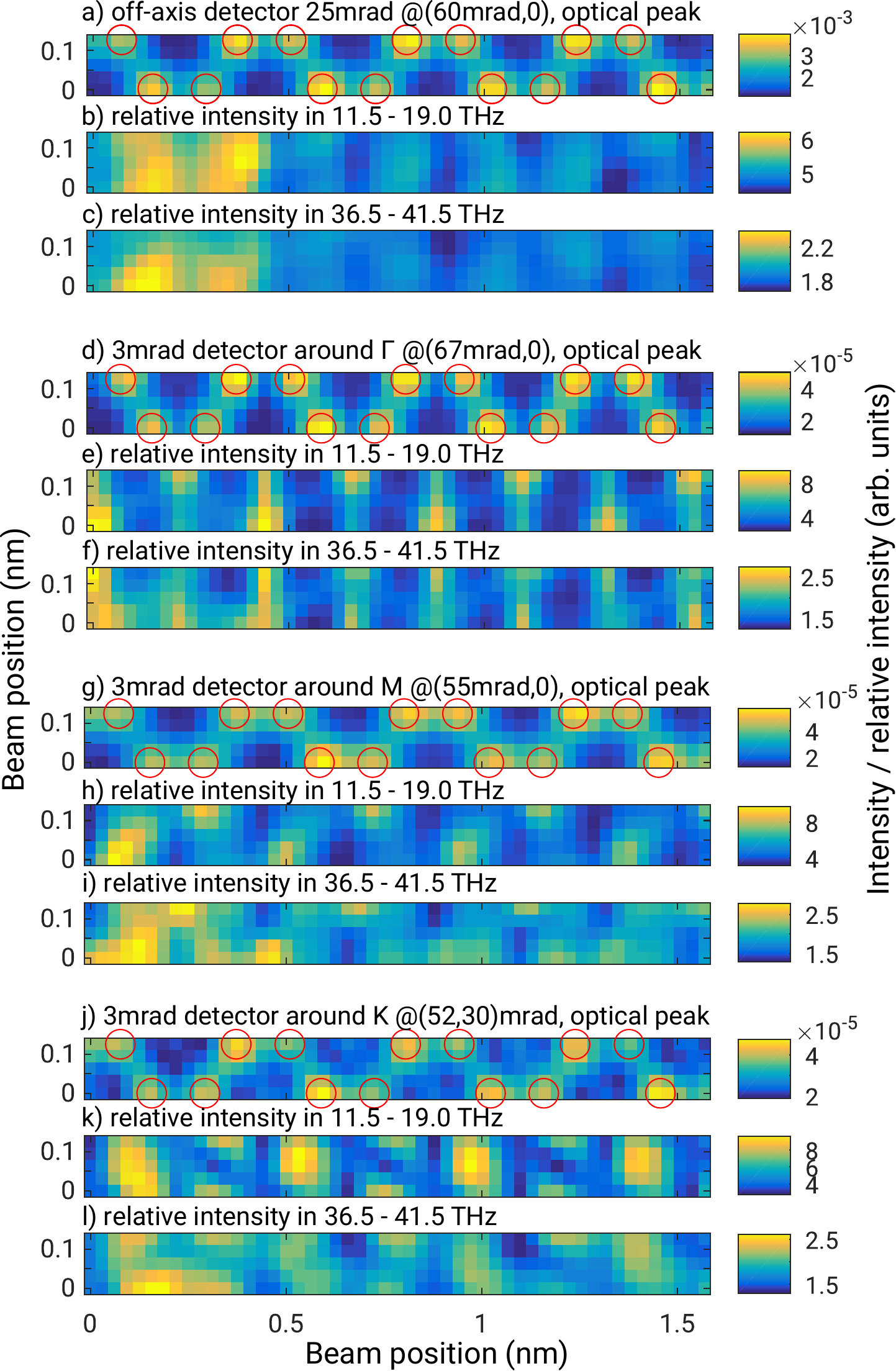}
    \caption{Collection of STEM-EELS images for four different detector settings: a)-c) large off-axis detector with 25~mrad collection semi-angle centered at $(60,0)$~mrad, d)-f) detector with 3~mrad collection semi-angle centered around $\Gamma$-point at $(67,0)$~mrad, g)-i) the same detector centered around $M$-point at $(55,0)$~mrad and j)-l) around $K$-point at $(52,30)$~mrad. Within each group, the three panels show (from top to bottom) the intensity at optical peak (46.5~THz frequency bin), relative intensity within the 11.5--19.0~THz range normalized to the optical peak, and relative intensity within the 36.5--41.5~THz range normalized to the optical peak. Positions of atomic columns are indicated by red circles in the panels showing optical peak intensities.}
    \label{fig:manyscans}
\end{figure}

Both nano-meter scale and sub-atomic scale spectral shape changes get highlighted in images from STEM-EELS simulations, focusing on intensities (or relative intensities) within specific frequency ranges. Figure~\ref{fig:manyscans} shows a collection of such STEM images for four different detector settings within three different energy ranges. For each of the four detector settings we show a STEM image formed from the intensity of the optical phonon peak at 46.5~THz, see panels a), d), g) and j). All of them display increased intensities nearby the positions of atomic columns (marked with red circles) and along the nearest-neighbor bonds between them. A slight reduction of intensities can be observed near the defect plane, although the difference is rather small, as could also be expected on the basis of Fig.~\ref{fig:spec30mrad}. Spectral shape differences become much more elucidated in STEM images showing relative intensities, normalized to the optical peak intensity---in analogy with, e.g., Fig.~\ref{fig:REdiag}b. Figure~\ref{fig:manyscans}b,c show such normalized STEM images for two energy intervals, namely for 11.5--19.0~THz in panel b) and 36.5--41.5~THz in panel c), respectively, assuming the large 25~mrad off-axis detector discussed above. These frequency intervals coincide with light-blue and rose shaded regions in Fig.~\ref{fig:spec3mrad}a. We observe that the relative intensity in those two frequency ranges is enhanced in the closest vicinity of the defect plane. Simultaneously we see the more subtle the sub-atomic scale relative intensity changes. However, one should note the colorbar range, which spans about 20\% of relative intensity variation. We have seen in Fig.~\ref{fig:REdiag3mrad} that relative intensity variation can be significantly larger at small collection angles.

Figure~\ref{fig:manyscans}, panels d)--l) show STEM images for a small collection semi-angle of 3~mrad, centered around selected special points in the diffraction plane, namely the $\Gamma$-point at $(67,0)$~mrad, $M$-point at $(55,0)$~mrad and $K$-point at $(52,30)$~mrad, respectively. The same frequency ranges have been highlighted. The optical phonon peak intensities (Fig.~\ref{fig:manyscans}d,g,j) drop by almost two orders of magnitude when compared to Fig.~\ref{fig:manyscans}a, in accordance with approximately 70 times lower angular coverage of the 3~mrad detector, when compared to a 25~mrad detector. Of key importance here are the relative intensity ranges, where the minimal and maximal intensity values differ by a factor of two or more, see Fig.~\ref{fig:manyscans}e,f for $\Gamma$-point, Fig.~\ref{fig:manyscans}h,i for $M$-point and Fig.~\ref{fig:manyscans}k,l for $K$-point, respectively. Not only we observe significant spectral shape variations at sub-atomic scale, the STEM images are also all qualitatively different. In other words, when using an electron beam of atomic size and a detector spans a collection angle much smaller than the convergence angle and angular dimensions of the Brillouin zone, then we can observe significant spectral shape modifications within both sub-\AA{}ngstr\"{o}m spatial shifts and few-mrad shifts of the detector center.

\section{Conclusions}\label{sec:conclusion}

We have computationally analyzed a bulk hexagonal BN system with defect planes -- anti-phase boundaries. Distinct differences in the local phonon density of states between the in-grain and at-the-defect regions were demonstrated. These differences correlate well with the simulated vibrational electron energy loss spectra. We have shown that nanometer-sized probe is sensitive enough to spatially resolve such spectral differences. Calculated angle-resolved spectra preserve a wealth of information about the phonon dispersions as well as their modifications due to the defect plane. These modifications are particularly pronounced at the scattering angles in the vicinity of $K$ and $M$ special points. Simulations with atomic size electron probe offer a detailed spatial picture about the spectral shape modifications. We observe larger scale modifications due to the presence of the defects as well as more subtle sub-atomic scale spectral shape variations. These variations become strongly enhanced in spectra simulated with a small collection angle.

Along with the detailed predictions of a vibrational-spectroscopic analysis of anti-phase boundaries in hexagonal BN for large scattering angles, from a more general perspective, our results demonstrate the properties of the frequency-resolved frozen phonon multislice method--reproducing the subtleties of recent experimental works at both nano-scale and atomic resolution, while being also computationally efficient, capable to deal with a structure model containing over 22000 atoms, thanks to its linear scaling with number of atoms.

\begin{acknowledgments}
This research is funded by the Swedish Research Council and Swedish National Infrastructure for computing (SNIC) at the NSC center (cluster Tetralith).
\end{acknowledgments}

\bibliography{references,references_special}

\end{document}